\documentclass[fleqn,usenatbib]{mnras}

\usepackage{newtxtext,newtxmath}

\usepackage[T1]{fontenc}
\usepackage{comment}
\DeclareRobustCommand{\VAN}[3]{#2}
\let\VANthebibliography\thebibliography
\def\thebibliography{\DeclareRobustCommand{\VAN}[3]{##3}\VANthebibliography}

\usepackage{graphicx}	
\usepackage{amsmath}	

\title[A reanalysis of the LHS 1140 b atmosphere]{A reanalysis of the LHS 1140 b atmosphere observed with the Hubble Space Telescope}

\author[Biagini A. et al.]{
Alfredo Biagini,$^{1,2}$\thanks{E-mail: alfredo.biagini@inaf.it}
Gianluca Cracchiolo,$^{1,2}$
Antonino Petralia,$^{2}$
Jes\'us  Maldonado,$^{2}$
Claudia Di Maio,$^{2}$\cr
Giuseppina Micela$^{2}$
\newauthor{}
\\
$^{1}$Università degli Studi di Palermo, Dipartimento di Fisica e Chimica, Via Archirafi 36, Palermo, Italy\\
$^{2}$INAF - Osservatorio Astronomico di Palermo, Piazza del Parlamento 1, 90134 Palermo, Italy
}

\date{Accepted March 15th, 2024. Received February 23rd, 2024; in original form December 15th, 2023}

\pubyear{2023}

\begin{document}
\label{firstpage}
\pagerange{\pageref{firstpage}--\pageref{lastpage}}
\maketitle

\begin{abstract}
The super-Earth LHS 1140 b is an interesting target for atmospheric observations since it is close to the habitable zone of its star and falls in the gap of the radius distribution of small exoplanets, in the region thought to correspond to the transition between planets with and without atmospheres. Observations of the primary transit with WFC3 on board of the Hubble Space Telescope (HST) revealed a modulation in the planet transmission spectrum compatible with the presence of water; however this modulation may be also due to stellar activity-related phenomena. Here we present a detailed analysis of the WFC3/HST observations to probe the nature of this modulation and to understand if it can be attributable to the presence of unocculted spots on the stellar surface. Our analysis strongly suggests that LHS1140 is a rather quiet star with subsolar [Fe/H] and enriched in $\mathrm \alpha$ elements. Therefore, we rule out the possibility that the planetary spectrum is affected by the presence of spots and faculae. This analysis shows the importance of a proper modelling of the stellar spectrum when analyzing transit observations. Finally, we modelled the planetary atmosphere of LHS1140 b to retrieve its atmospheric composition. However, the low resolution and the narrow spectral range of HST observations prevented us from definitively determining whether the spectral features are attributable to the presence of water or of other molecules in the planetary atmosphere.

\end{abstract}




\section{Introduction}
Nowadays the study of exoplanets and their atmosphere is a leading topic of astronomical research. The main method to study this kind of targets is the transit method, both using photometry and spectroscopy. Specifically, in-transit spectroscopic observations are fundamental to study the atmosphere of exoplanets.
The extraction of the planetary spectrum from the transit observations is based on the assumption of a uniform and quiet stellar surface, thus reducing the transit to a purely geometrical effect, but this is not the case for active stars. 
In fact, the stellar surface of active stars presents active regions which bias the correct derivation of both stellar and planetary parameters, including convection-related phenomena, such as granulation and pulsations \citep[e.g.][]{2005LRSP....2....8B}, and magnetic activity-related phenomena, such as spots \citep{2003A&ARv..11..153S,2012A&A...539A.140B}, faculae and flares \citep[e.g.][]{2012A&A...539A.140B}.
Due to their lower temperature, star spots have a spectrum distinct from that of the surrounding photosphere, with greater contrast at shorter wavelengths. This chromatic dependence changes the stellar spectrum and may distort the entire transmission spectrum of the transiting planet if the distortion is large enough, and may also mimic a planetary atmosphere even if there is none. 
Therefore, the presence of spots may hamper the extraction of the final transmission spectrum of the planetary atmosphere, for both photometric and spectroscopic observations \citep[e.g.][]{2008MNRAS.385..109P,2009A&A...505.1277C,2009A&A...505..891S,2011MNRAS.416.1443S,2011A&A...527A..73S,2010ApJ...721.1861A,2011ApJ...736...12B,2011A&A...526A..12D,2012A&A...539A.140B,2015ExA....40..723M,2015ExA....40..711S}. The existence of activity-related “pseudo”-transmission effects is well known and has been discussed, e.g., in \cite{2018A&A...620A..97S}, in the context of high-resolution atomic lines. They show that stellar activity related pseudo-signals mix with and confuse the planetary atmospheric absorption signal. Without an accurate analysis of stellar activity the atmospheric analysis remains often uncertain.
A number of methods to extract the planetary spectrum taking into account the stellar activity have been proposed in literature \citep{2009A&A...505.1277C,2011A&A...527A..73S,2012A&A...539A.140B,2015ExA....40..723M,2021MNRAS.501.1733C,2021MNRAS.507.6118C,2023arXiv230204574T}.

On the other hand, the stellar spectrum depends on the stellar metallicity (Z) and the rate of $\mathrm \alpha$ elements (C, O, Ne, Mg, Si, S, Ar and Ca) in the stellar atmosphere can affect deeply the stellar spectrum leading to a wrong interpretation of the planetary spectrum if not correctly accounted for.
So, anticipating one of the results of this work, a proper modelling of the stellar spectrum is mandatory in order to properly characterize the planetary atmosphere. 

Today, many space missions, currently in flight like JWST \citep[James Webb Space Telescope,][]{2016ApJ...817...17G,2006SSRv..123..485G} or in the development stage, e.g. ARIEL \citep[Atmospheric Remote-Sensing Infrared Exoplanet Large-survey,][]{2018ExA....46..135T}, aim at observing the extrasolar planets to understand their chemical and physical properties through their atmospheres; therefore, in order to achieve this goal, we need to improve our ability to remove chromatic distortions caused by the star.\\

In this work we analyzed the case of LHS 1140 b to understand how our knowledge of the host star might impact the extraction of the planetary signal and, hence, the retrieval of its atmospheric composition. 
With an equilibrium temperature of about $\mathrm{  230\: K}$, LHS 1140 b is situated close to the habitable zone (HZ) of its star \citep{2017Natur.544..333D,2018ApJ...861L..21K,2023arXiv231015490C}, where the existence of liquid water on the planet surface is possible. Recent ground-based observations are not precise enough to constrain atmospheric scenarios \citep{2020AJ....160...27D}. However,  \cite{2021AJ....161...44E} using transit observations from Hubble Space Telescope (HST) Wide Field Camera 3 (WFC3) found a modulation in the transit depth of LHS 1140 b over the $\mathrm{ 1.1-1.7 \: \mu m}$ wavelength range. The authors investigated this modulation, both evaluating multiple atmospheric scenarios to fit the data (trace of $ \mathrm{H_{2}O}$ in the atmosphere, a moist atmosphere and an atmosphere with $ \mathrm{CH_{4}}$ and also enriched with $ \mathrm{N_{2}}$) and investigating the possibility of contamination from stellar activity, specifically due to the presence of stellar spots. \\
In this paper, we aim to further evaluate the spectral contamination due to the stellar activity adopting the approach of \cite{2021MNRAS.507.6118C} in order to better understand the nature of this system. We analyzed the WFC3/HST observations of LHS 1140 b to understand if this modulation is attributable to the presence of water in the planetary atmosphere or is an artifact due to the presence of spots on the stellar surface.

This paper is organized as follows: in Section 2 we shortly describe the LHS 1140 system and in Section 3 we explain the reduction pipeline that we used to analyze HST data. In Section 4 we derived the temperature and composition of the star from the out-of-transit observation. Section 5 investigates the possibility of stellar activity on this star, while Section 6 explores a discussion on its potential stellar population. Finally, in Section 7, we examine the implications of the stellar composition on the study of LHS 1140b.

\section{LHS 1140 \MakeLowercase{b}}

LHS 1140 is a M4.5-type star \citep[m$_{\rm V}$   = 14.5][]{2013AJ....145...44Z}, located in the costellation of the Whale, at a distance of at ($\mathrm{12.47 \pm 0.42}$) pc \citep{2017Natur.544..333D,2023A&A...674A...1G} from the solar system.
It has a surface temperature of $\mathrm{ 3096 \pm 48}$ K \citep{2023arXiv231015490C} while \cite{2019AJ....157...32M} estimated a value of $\mathrm{ 3216 \pm 39}$ K. The star has a mass of $\mathrm{  0.184 \pm 0.005 \: M\odot }$ and a radius of $\mathrm{  0.216 \pm 0.003 \: R\odot}$ \citep{2023arXiv231015490C}.\\
In April 2017, the exoplanet LHS 1140 b was discovered, orbiting  at $\mathrm{ 0.0946 \pm 0.0017 \: AU}$ from its star with a period of $\mathrm{ 24.73723 \pm 0.00002 \:days}$ \citep{2020MNRAS.494.5082G,2023arXiv231015490C}. Its radius is $\mathrm{  1.730 \pm 0.025}$ $\mathrm{ R_{\oplus}}$, as derived by \cite[][and references therein]{2019AJ....157...32M,2023arXiv231015490C} from photometric transit observations of Spitzer \citep{2004ApJS..154....1W} and MEarth-South \citep{2012AJ....144..145B} survey. Using radial velocity measurements from the high-resolution HARPS spectrograph \citep{2003Msngr.114...20M},  \cite{2023arXiv231015490C} estimated a planetary mass $\mathrm{ M_{p}= 5.60 \pm 0.19}$ $\mathrm{  M_{\oplus}} $, resulting in a density of $ \mathrm{ 5.9 \pm 0.3}$ $\mathrm{g\:cm^{-3}} $, very similar to the Earth density. Its equilibrium temperature is evaluated as $\mathrm{226 \pm 4 \: K}$ \citep{2023arXiv231015490C}. \\

We analyzed HST data from two transit observations of LHS 1140 b that were acquired for proposal 14888 (PI: Jason Dittmann) with the WFC3/HST G141 grism and were taken in January and December 2017. Our analysis starts from the raw spatially scanned spectroscopic images collected with Hubble WFC3 and obtained from the Mikulski Archive for Space Telescopes (MAST)\footnote{\url{https://mast.stsci.edu/portal/Mashup/Clients/Mast/Portal.html }}.
Both visits utilised the GRISM256 aperture, and 256 × 256 subarray, with an exposure time of 103.13 s which consisted of 16 up-the-ramp non-destructive reads using the SPARS10 sequence. The visits had different scan rates with 0.10"/s and 0.14"/s used for January and December respectively, resulting in scan lengths of 10.9" and 15.9".
In 2018, a second terrestrial planet orbiting LHS 1140 was discovered with the primary transit technique: LHS 1140 c \citep{2019AJ....157...32M}, orbiting at $\mathrm{0.0270  \pm 0.0005 \: AU}$ from its star, with an orbital period of $\mathrm{3.777940 \pm 0.000002 \: days}$ and an equilibrium temperature of $\mathrm{438 \pm 9}$ $\mathrm K$ \citep{2023arXiv231015490C}. Its radius is $\mathrm{1.272 \pm 0.026 \: R_{\odot}}$ while its mass is $\mathrm{1.91 \pm 0.06}$, with a mean density of $\mathrm{ 5.1 \pm 0.4}$ $\mathrm{g\:cm^{-3}}$ \citep{2023arXiv231015490C}.\\

LHS 1140 b and c are ideal targets for atmospheric characterization with  the JWST and the upcoming Ariel mission: they are both super-Earths orbiting an M dwarf star and the planet b is potentially in its HZ. Moreover \cite{2019A&A...624A..49W} showed through simulations that $\mathrm{H_{2}O}$, $\mathrm{CH_{4}}$ and also $\mathrm{CO_{2}}$ should be detectable in the case of LHS 1140 b.

\begin{table}
\centering
\caption{Stellar parameters of LHS 1140 and planetary parameters of LHS 1140 b  taken from \citet{2023arXiv231015490C}. The value of the mid-transit time, T$_{\rm mid}$, is derived in this work.}
\label{PAR}
\begin{tabular}{| c | c |}
\hline
\hline
\textbf{PARAMETER} &\textbf{VALUE}\\
\hline
\rule{0pt}{4ex}$\mathrm{ R_{\star}} [\mathrm R_{\odot}]$& $\mathrm  0.216 \pm 0.003$\\
\rule{0pt}{4ex}$\mathrm{ M_{\star}} [\mathrm M_{\odot}]$& $\mathrm  0.184 \pm 0.005 $\\
\rule{0pt}{4ex}$\mathrm{ T_{\star}} [\mathrm K]$& $\mathrm 3096 \pm 48$\\
\rule{0pt}{4ex}$\mathrm{ log (g)}$& $\mathrm 5.041 \pm 0.016$\\
\hline
\rule{0pt}{4ex}$\mathrm{ M_{p}} [\mathrm M_{\oplus}]$& $\mathrm 5.60 \pm 0.19$\\
\rule{0pt}{4ex}$\mathrm{ R_{p} [R_{\oplus}]} $& $\mathrm 1.730 \pm 0.025$\\
\rule{0pt}{4ex}$\mathrm a [\mathrm{ AU}]$& $\mathrm  0.0946 \pm 0.0017$\\  
\rule{0pt}{4ex}$\mathrm{ i [deg]}$& $\mathrm 89.86 \pm 0.04$\\    
\rule{0pt}{4ex}$\mathrm e$& $ < 0.043 (95\%)$\\
\rule{0pt}{4ex}$\mathrm{ P_{orb} [d]}$& $\mathrm  24.73723 \pm 0.00002$\\
\rule{0pt}{4ex}$\mathrm{ T_{mid} [ BJD_{TDB}]}$& $\mathrm  2458103.08346 \pm 0.00005$\\
\hline

\end{tabular}
\end{table}

\begin{figure}
		\centering
            \includegraphics[width=\columnwidth]{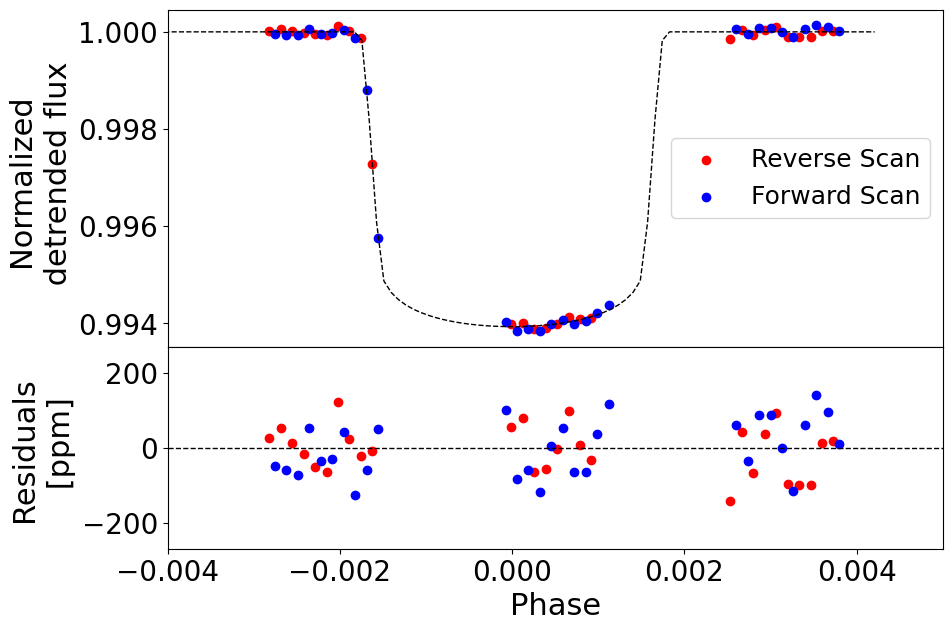}
		\caption{\label{FIT} White light curve fit for December observation of LHS 1140 b. Top: detrended flux and best-fit model. Bottom: residuals from best-fit model.}
\end{figure}%

\begin{figure*}
		\centering
		\includegraphics[width=\linewidth]{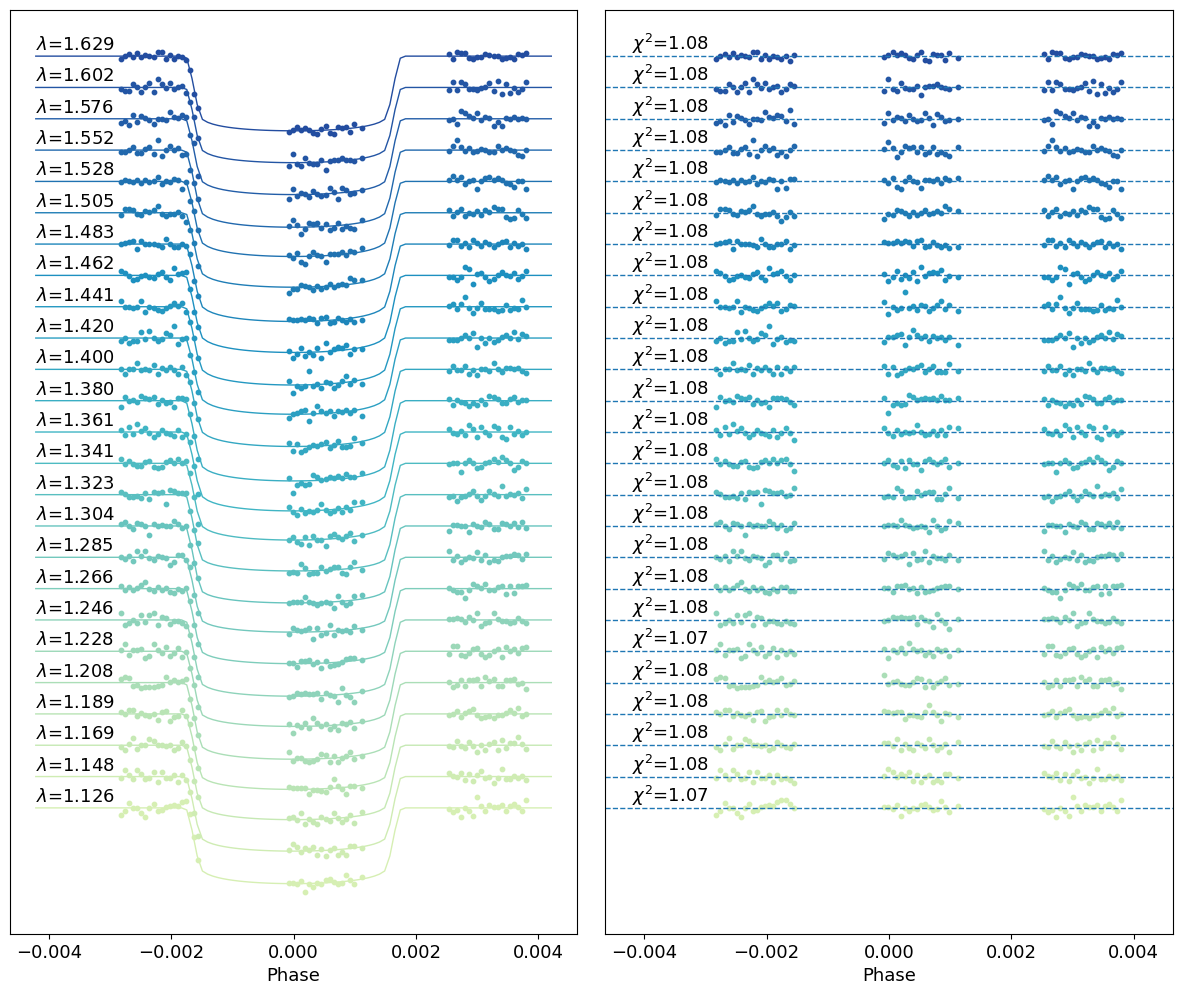}
		\caption{\label{SP_LC} Normalized spectral light curve fits from Iraclis for the transmission spectrum of LHS 1140 b where, for clarity, an offset has been applied. Left panel: the detrended spectral light curves with the best-fit models plotted; right panel: residuals from the fitting to the reported values of the Chi-squared ($\mathrm{\chi^{ \rm 2}}$).}
\end{figure*}%

\section{REDUCTION AND ANALYSIS OF WFC3/HST DATA}
Before we proceed further in the analysis of the HST data, it is mandatory to briefly summarize the main steps involved in the reduction of the data. 
Although two transits of LHS 1140 b were obtained using WFC3, the January visit was discarded from our analysis since it was affected by large shifts in the location of the spectrum on the detector  \citep[see also][]{2021AJ....161...44E}. For this reason we focus only on the December observation, which uses both forward and reverse scanning modes of the instrument.
In the following, Iraclis\footnote{\url{https://github.com/ucl-exoplanets/Iraclis} } \citep{2018AJ....155..156T}, a specialised open-source software for the analysis of WFC3 scanning observations, was used. The reduction process includes the following steps: zero-read subtraction, reference pixels correction, non-linearity correction, dark current subtraction, gain conversion, sky background subtraction, flat-field correction, and corrections for bad pixels and cosmic rays \citep[for a detailed description of these steps, see the original Iraclis paper,][]{2018AJ....155..156T} and this reduction pipeline is common for both out-of-transit and in-transit spectra.  For the December observation, the reduced spatially scanned spectroscopic images were used to extract the white (from $\mathrm{1.1-1.7 \: \mu m}$) and spectral light curves. The spectral light curves bands were selected as in \cite{2021AJ....161...44E}, that is in such a way that the signal-to-noise ratio (SNR) is approximately uniform across the planetary spectrum. The first orbit of each visit was discarded as it presents stronger wavelength-dependent ramps, and also the first exposure in each HST orbit after each buffer dump was removed since it contains significantly lower counts than subsequent exposures \citep[e.g.][]{2013ApJ...774...95D,2016ApJ...832..202T}. The light curves were fitted by using the transit model package PyLightcurve \citep{2016ApJ...832..202T} which utilises the MCMC (Markov Chain Monte Carlo) code emcee  \citep{2013PASP..125..306F}. The only free parameters for the fitting of the white light curve were the mid-transit time and the planet-to-star radius ratio. The other planet parameters were fixed to the values from \cite{2023arXiv231015490C} ($\mathrm{a/R_{\star}= 94.47,  i = 89.86^{\circ}}$)while we used the same limb darkening coefficients (LDCs) of \cite{2021AJ....161...44E}. The stellar parameters used to download the LDCs were taken from \cite{2021AJ....161...44E} and the selected database was PHOENIX-2012-13\footnote{\url{https://phoenix.ens-lyon.fr/Grids/}} \citep{2012A&A...546A..14C}. The WFC3 exoplanet observations are usually affected by two kinds of time-dependent systematics: the long-term and short-term “ramps” \citep{2018AJ....155..156T}. In Iraclis, these systematics in the white time series are fitted by using Eq. \eqref{EQ1}:

\begin{equation}
     \mathrm{ R_{w}(t)  = n^{scan}_{w}  [ 1- r_{a} (t - T_{0})] (1 - r_{b_{1}} e^{-r_{b_{2}} (t - t_{0})})}
	\label{EQ1}
\end{equation}

where t is time, $\mathrm{ n^{scan}_{w}}$  is a normalisation factor, $\mathrm{T_{0}}$ is the mid-transit time, $\mathrm{ t_{0}}$ is the time when each HST orbit starts, $\mathrm{ r_{a}}$ is the slope of a linear systematic trend along each HST visit and $\mathrm{r_{b_{1}} ,r_{b_{2}}}$ are the coefficients of an exponential systematic trend along each HST orbit. The normalisation factor $\mathrm{ n^{scan}_{w}}$ is the averaged out-of-transit flux in the white light curve. $\mathrm{ n^{scan}_{w}}$ is adapted to $\mathrm{n^{for}_{w}}$  for forward scanning mode and to $\mathrm{ n^{rev}_{w}}$ for reverse scanning mode. The reason for using different normalisation factors is the slightly different effective exposure time due to the known upstream/downstream effect \citep{2012wfc..rept....8M}. Initially, the white light curve is fitted by using Eq. \ref{EQ1} with Iraclis. Next, the spectral light curves are fitted with a transit model where the only free parameter is the planet-to-star radius ratio along with a model for the systematics ($\mathrm{ R_{\lambda}}$) that included the white light curve \citep[divide-white method,][]{2014Natur.505...69K} and a wavelength-dependent, visit-long slope \citep{2018AJ....155..156T} parameterised by Eq. \eqref{EQ2}: \\
\begin{equation}
     \mathrm{ R_{\lambda}(t) = n^{scan}_{w} (1-\chi_{\lambda}(t-T_{0})) \dfrac{LC_{w}}{M_{w}}}
	\label{EQ2}
\end{equation}

where $\mathrm{ \chi_{\lambda}}$ is the slope of a wavelength-dependent linear systematic trend along each HST visit, $\mathrm{ LC_{w}}$ is the white light curve and $\mathrm M_{w}$
is the best-fit model for the white light curve. The normalisation factor $\mathrm{ n^{scan}_{w}}$ is the averaged out-of-transit flux in each spectral bin centered at $\lambda$. Again, the normalisation factor $\mathrm{ n^{scan}_{w}}$ is changed to $\mathrm{ n^{for}_{w}}$ for upward scanning directions (forward scanning) and to $\mathrm{ n^{rev}_{w}}$ for downward scanning directions (reverse scanning). The white light curve fit is shown in Figure \ref{FIT} and the subsequent spectral light-curve fits are shown in Figure \ref{SP_LC}. A full list of stellar and planet parameters used here for the fitting is given in Table \ref{PAR}.
It should be noted that in this work we use the out-of-transit spectrum of the star to model its properties. In principle there could be some chromatic time-dependent systematics in the absolute flux estimations that could affect the retrieval of the stellar properties based on the out-of-transit spectrum of the star. We verified that these systematics, if present, are negligible for our analysis as it is shown it in APPENDIX A. It should also be noted that Iraclis corrections does not correct possible residual wavelength-dependent systematics in the absolute flux.

\begin{figure*}
    \centering

      \includegraphics[width=\columnwidth]{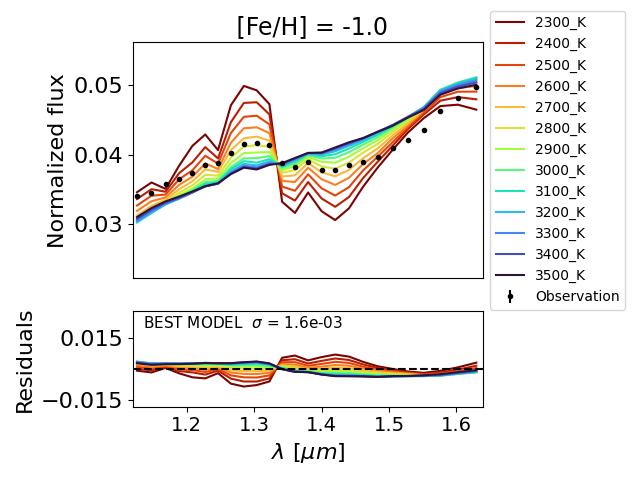}
      \includegraphics[width=\columnwidth]{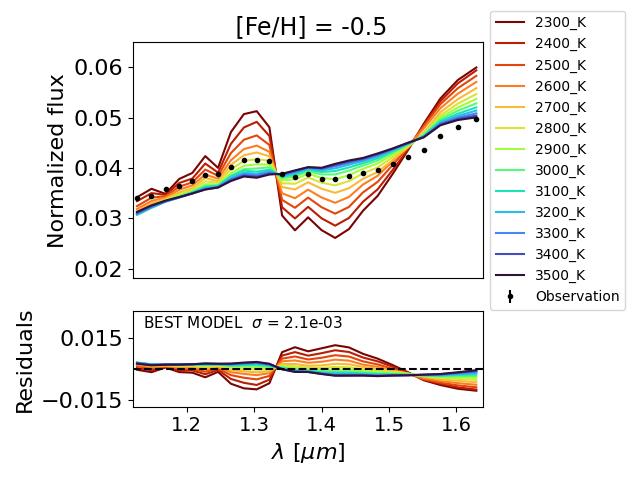}

      \includegraphics[width=\columnwidth]{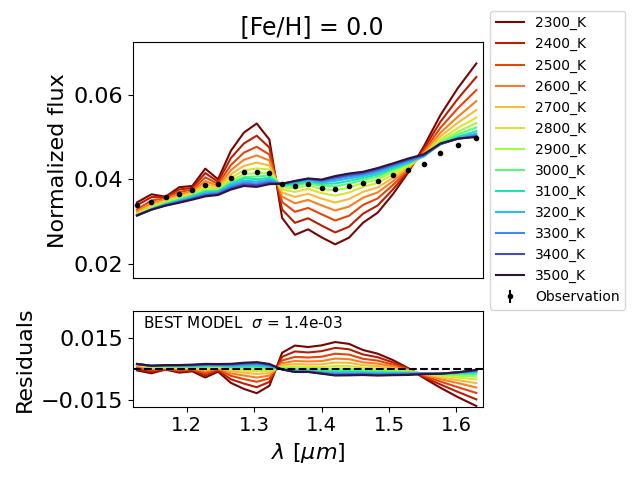}
      \includegraphics[width=\columnwidth]{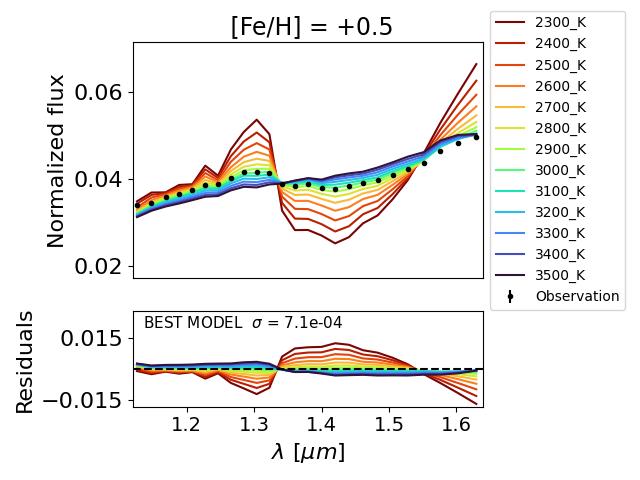}

     \centering
     \includegraphics[width=\columnwidth]{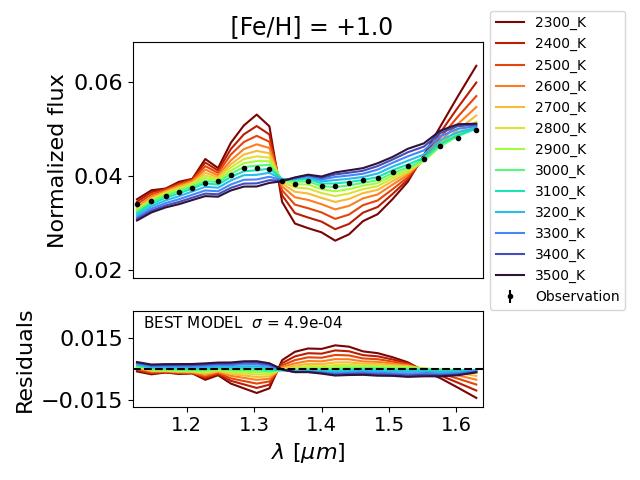}
   
    \caption{\label{Met1} Comparison between the observed out-of-transit stellar spectrum (dots, uncertainties are within the symbol size) and the simulated Phoenix spectra at different temperatures (2300-3500 K, see legends) and at different compositions: [Fe/H] = -1.0 dex (top left),  [Fe/H] = -0.5 dex (top right), [Fe/H] = 0.0 dex (center left),  [Fe/H] = +0.5 dex (center right),  [Fe/H] = +1.0 dex (bottom). In each graph we show the standard deviation of the best model for each given stellar metallicity.} 
\end{figure*}

\begin{figure*}
		\centering
                \includegraphics[width=\columnwidth]{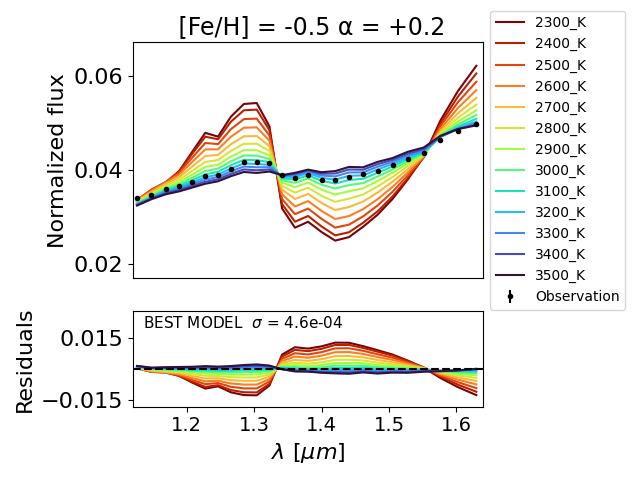}
		\includegraphics[width=\columnwidth]{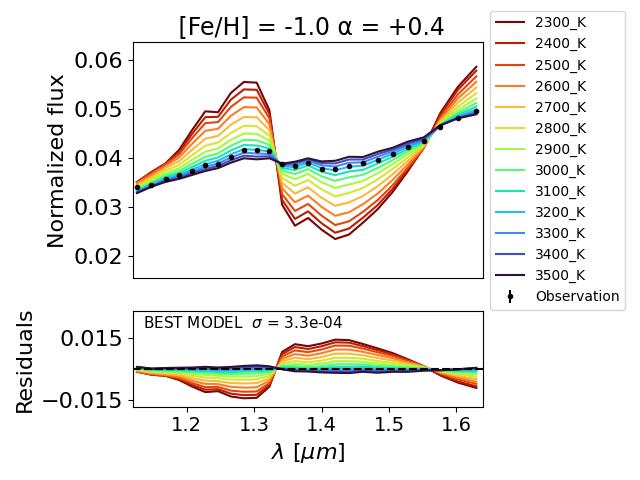}
		\caption{\label{Met2} Comparison between the observed out-of-transit stellar spectrum (dots, uncertainties are within the symbol size)  and simulated spectra corresponding to an $\mathrm \alpha$ enriched star: the left panel corresponds to a composition of $\mathrm{ [Fe/H] = -0.5}$ $\mathrm{ dex}$ and $\mathrm \alpha = +0.2$ and the right panel corresponds to a composition of $\mathrm{[Fe/H] = -1.0}$ $\mathrm{ dex}$ and $\mathrm{\alpha = +0.4}$, both cases with a range of temperatures between 2300 and 3500 K. In each graph we show the standard deviation of the best model for each given stellar composition.}
\end{figure*}%

\begin{figure*}
		\centering
		\includegraphics[width=\columnwidth]{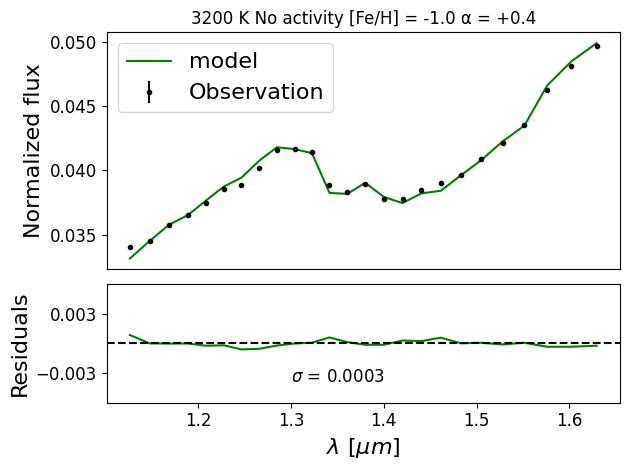}
        \includegraphics[width=\columnwidth]{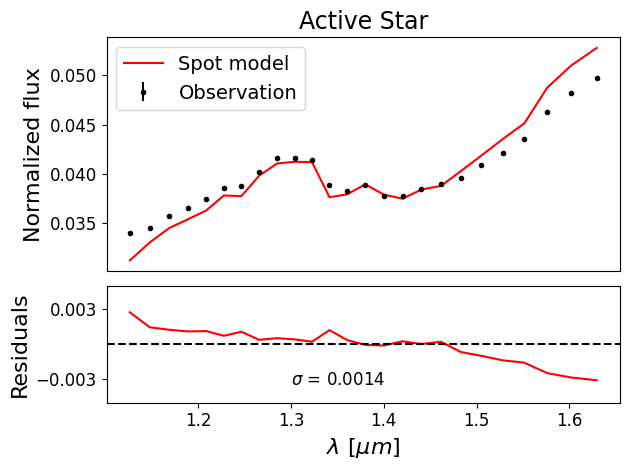}
		\caption{\label{Good_fit} Left side: comparison between the observed HST data (black dots) and the simulated phoenix spectrum at 3200 K and with a composition of  [Fe/H] = -1.0 and $\mathrm{ \alpha}$ = +0.4 (green line). Right side: comparison between the same data (black dots) with the best-fit spectrum (red line) corresponding to the best one spot model found by our stellar activity simulations. In both images residuals are shown in the lower panel with the related standard deviation.} 
\end{figure*}%

\begin{figure*}
		\centering
		\includegraphics[scale=0.5]{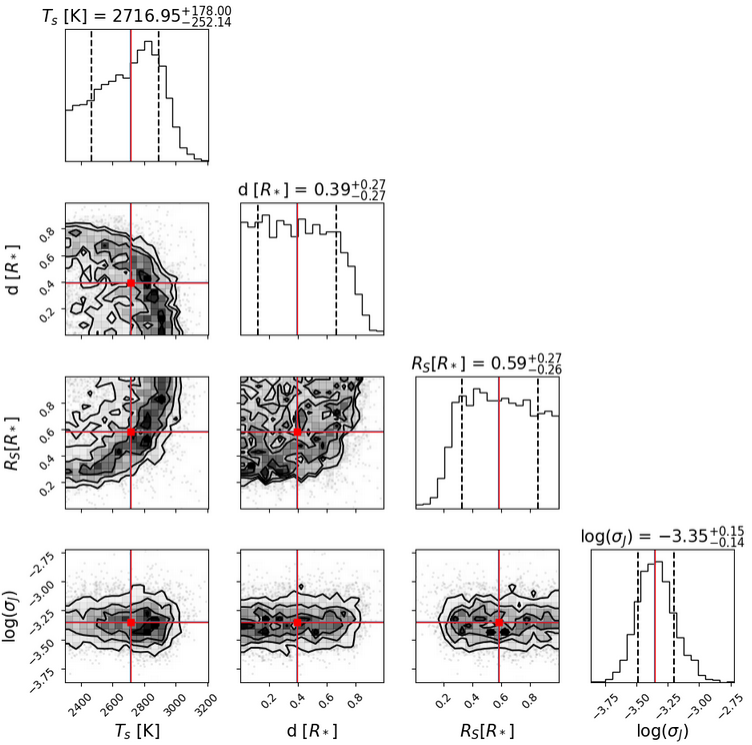}
		\caption{\label{CORNER_SPOT} Corner plot of the best-fit parameters obtained from the fitting of $\mathrm{n^{scan}}$ in forward model using a one spot model for stellar activity simulations. $\mathrm{T_{s}}$ is the temperature of the spot, d is the distance between the center of the spot and the center of the stellar disk. $\mathrm{R_{s}} $ is the radius of the spot normalized to the stellar radius $\mathrm{R_{\star}}$ and $\mathrm{\sigma_{j}}$ is the jitter noise. The red lines correspond to the values of maximum probability (MAP) for the fit and the dotted lines delimit their confidence interval.}
\end{figure*}%

\begin{figure}
		\centering
		\includegraphics[width=\columnwidth]{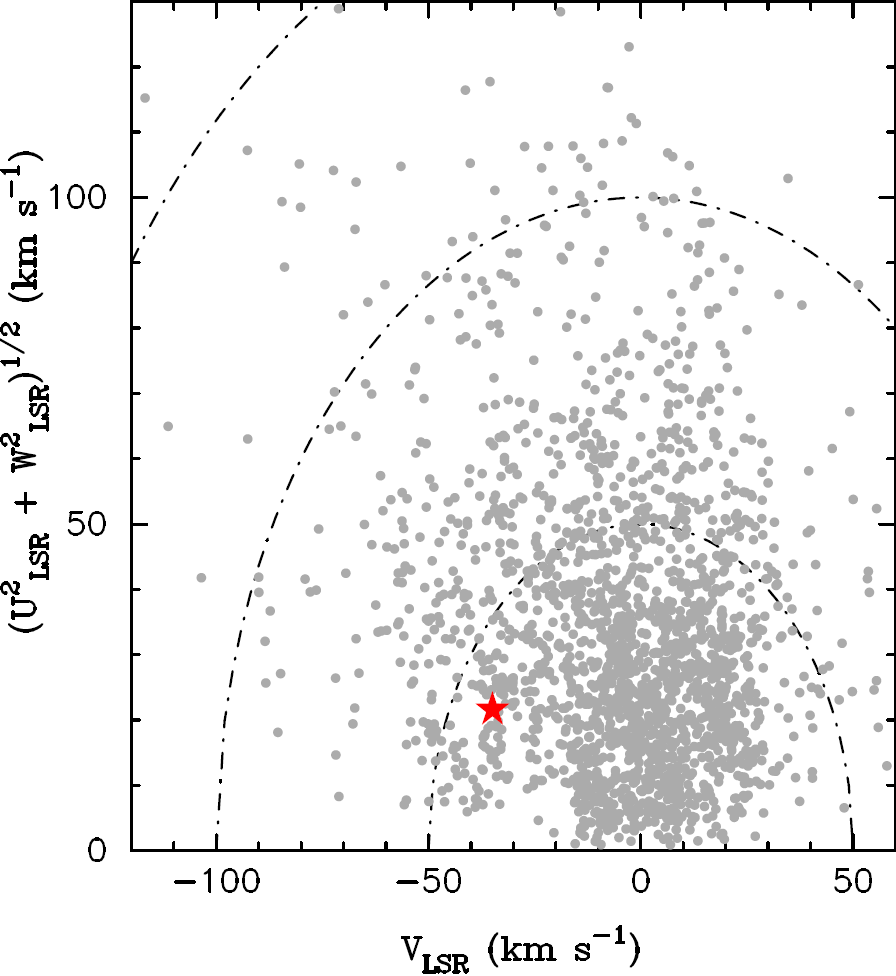}
		\caption{\label{Toomre_LHS.png} Toomre diagram for the stellar neighboorhood of LHS 1140 (shown as a red symbol in image). Dotted lines indicate constant peculiar space velocities, $\mathrm{v_{pec}} = (U_\mathrm{{LSR}}^{2} + V_\mathrm{{LSR}^2} + W\mathrm{_{LSR}^{2})^{1/2}}$ in steps of 50  $\mathrm{ km \: s^{-1}}$, where $\mathrm{U_{LSR}}$, $\mathrm{V_{LSR}}$ and $W\mathrm{_{LSR}}$ are the velocities of the star with respect to the local standard of rest (LSR).   }
\end{figure}%

\begin{figure*}
		\centering
		
            \includegraphics[scale=0.7]{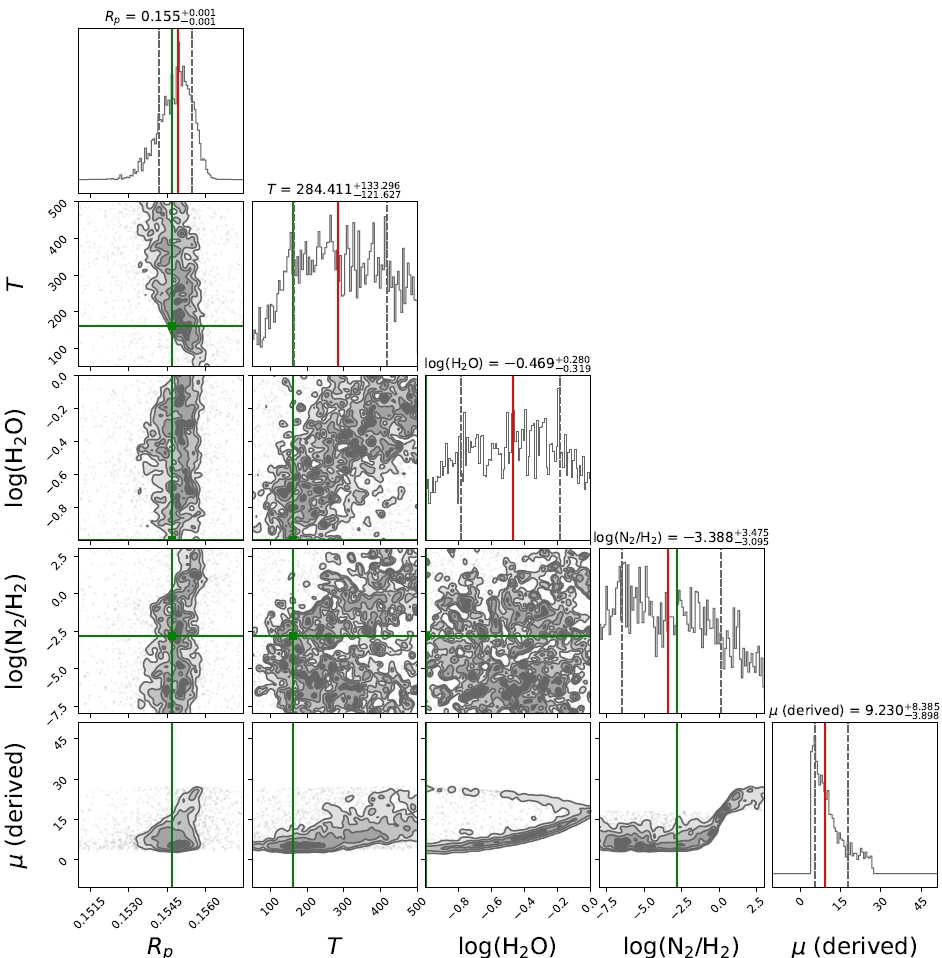}
		\caption{\label{POSTERIOR} Retrieval results obtained for best atmospheric model, corresponding to an atmosphere with $\mathrm{N_{2}}$, $\mathrm{H_{2}O}$ and without clouds. The green and red vertical solid lines highlight the maximum a posterior (MAP) and median values, respectively, while the vertical dashed-lines represent the values at 1$\mathrm \sigma$ from the median. }
\end{figure*}%

\begin{figure}
		\centering
		\includegraphics[width=\columnwidth]{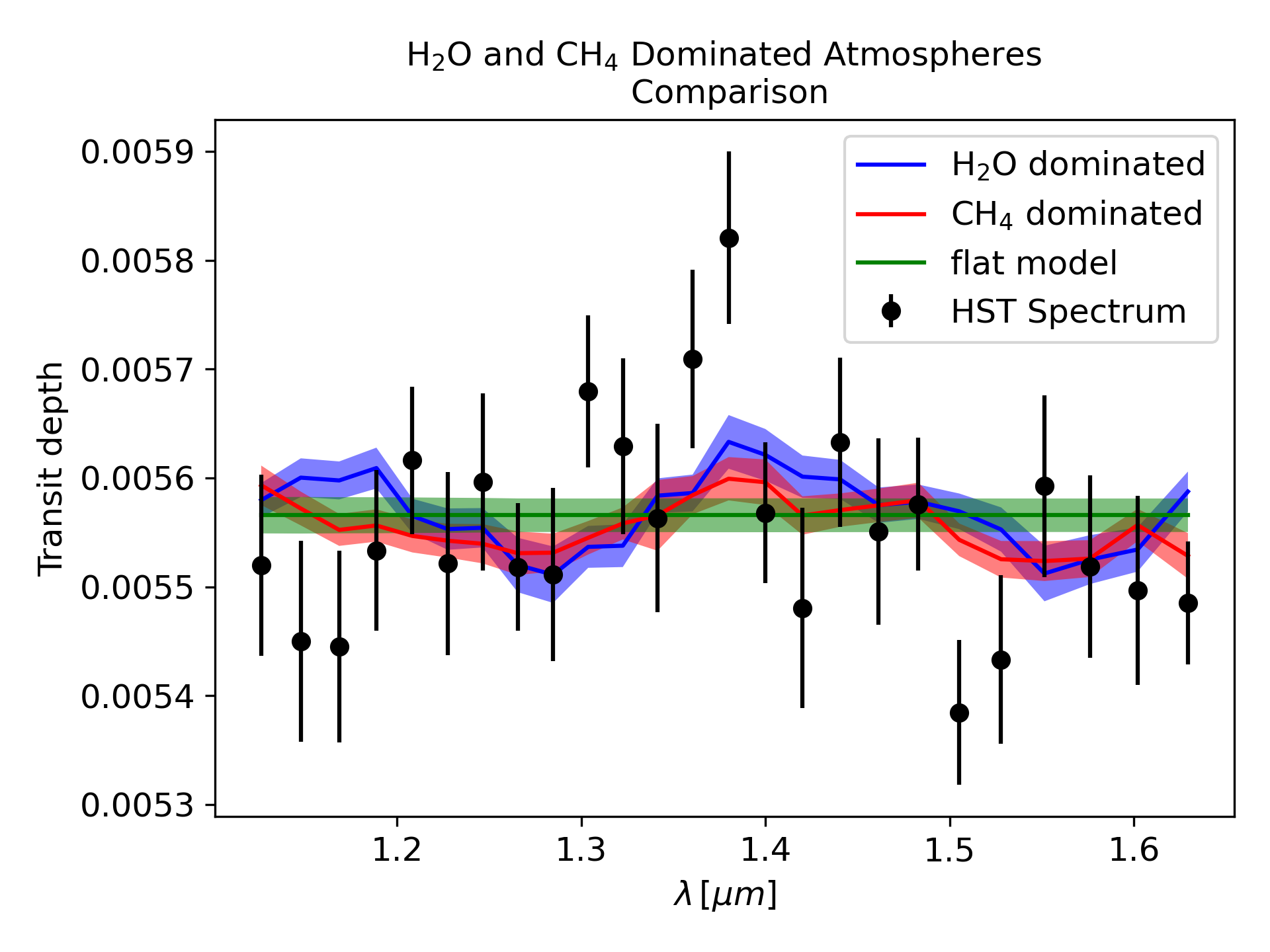}
		\caption{\label{DOMINATED} Comparison between the observed planetary spectrum (black dots) and the best fit for a secondary atmosphere dominated by $\mathrm{H_{2}O}$ (blue) and one dominated by $\mathrm{CH_{4}}$ (red), without clouds. The green line represents the flat model.}
\end{figure}%

\begin{figure}
		\centering
		\includegraphics[width=\columnwidth]{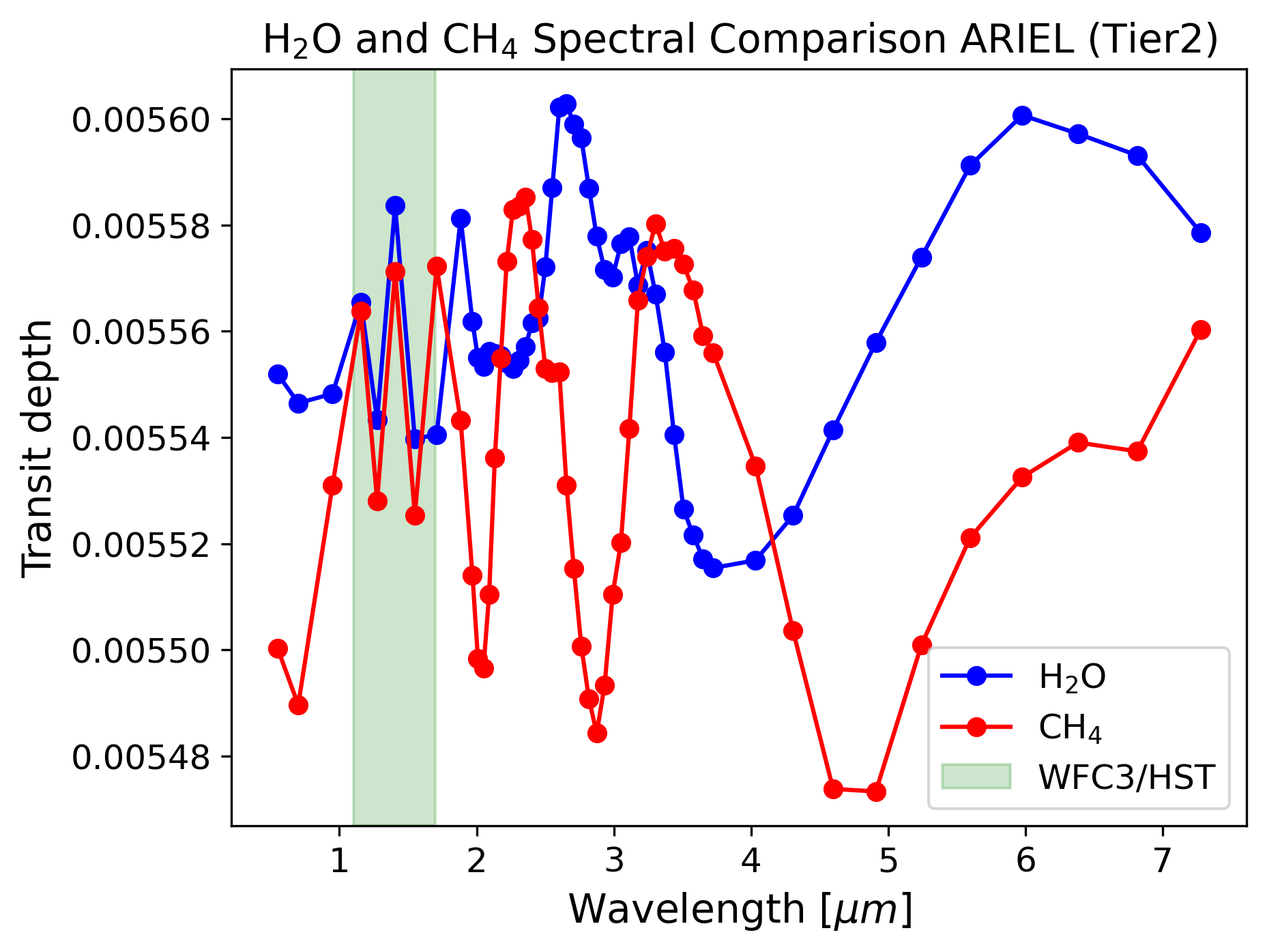}
            \includegraphics[width=\columnwidth]{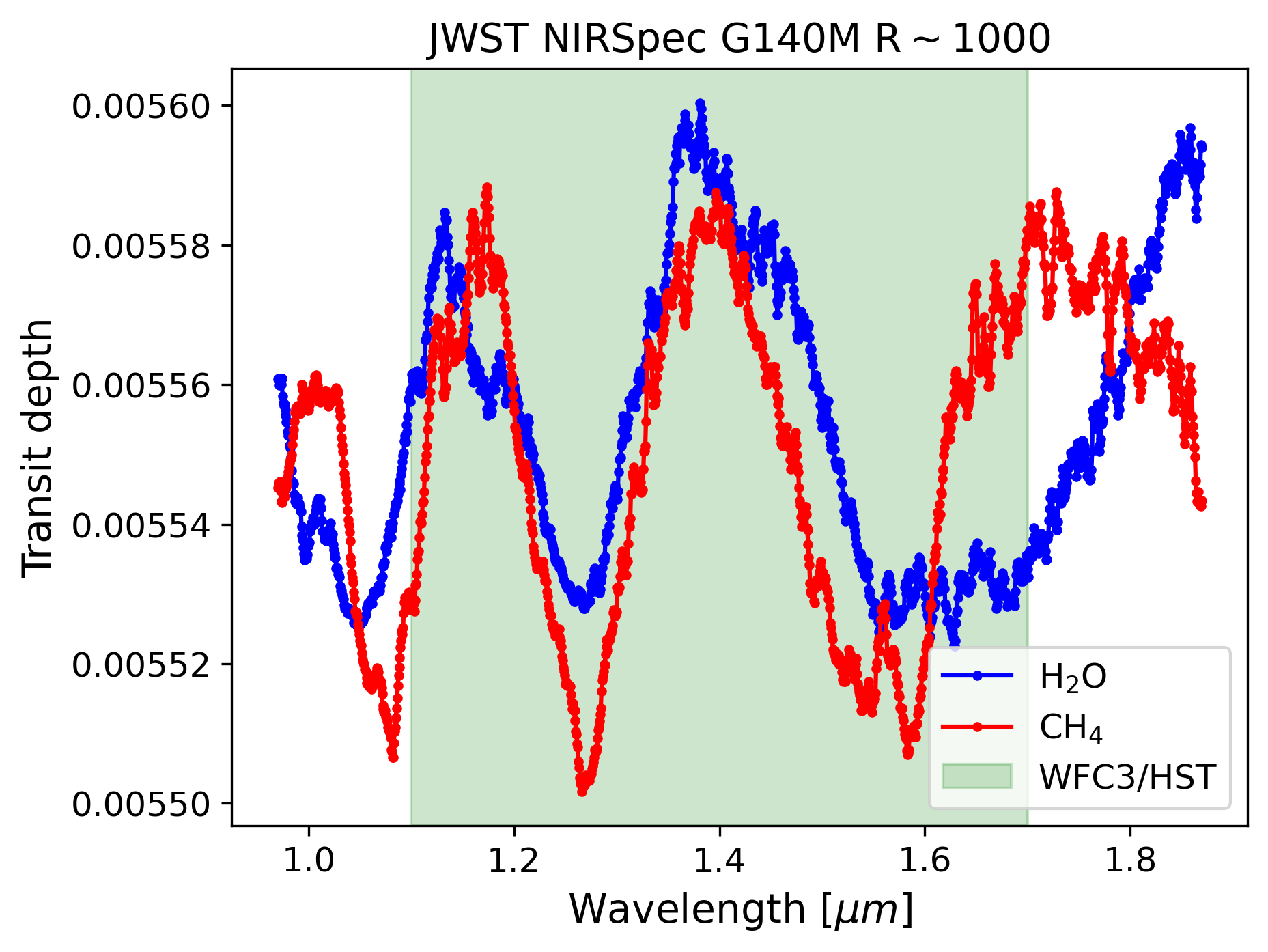}
		\caption{\label{ARIEL} Comparison between the expected transit spectrum of LHS 1140 b with an atmosphere enriched in $\mathrm{H_{2}O}$ or $\mathrm{CH_{4}}$, the same scenarios shown in Figure \ref{DOMINATED} as observed by ARIEL tier 2 (top) and JWST Nirspec with medium resolution G140M grism (bottom). The spectral range of WFC3/HST is shown by the green area.} 
\end{figure}%
\section{Spectral Analysis}

\cite{2021AJ....161...44E} suggested the possibility that some features of the spectrum of LHS1140 b could be due to stellar spots contaminations and tried to simulate them. However, the X-ray luminosity value  ($\mathrm{ L_{x}}$) of LHS 1140 b as reported by \cite{2023AJ....165..200S} put this star at the lower end of the X-ray distribution function ($\mathrm{ L_{x} \sim 1.4 \times 10^{26}}$ $\mathrm{ erg \:s^{-1}}$) pointing to a quiet old star. Moreover, the analysis of the TESS lightcurves shows no significant signs of activity \citep{2022AAS...24030903B}. 
LHS 1140 has also a long rotational period of about 131 days \citep{2017Natur.544..333D,2018AJ....156..217N} suggesting a low level of activity \citep{2003A&A...397..147P}. 
These observational evidences suggest that LHS 1140 is rather a quiet, inactive star. Therefore, it is rather unlikely that stellar spots might severely affect its stellar spectrum. In the following, we will test the alternative hypothesis that the star has a non-solar chemical composition.

\subsection{Stellar Composition}
In order to do so, we evaluated various stellar atmospheric models to determine which one could best describe the observed out-of-transit spectrum. 
We generated a set of stellar spectra obtained from the PHOENIX-2012-2013 database to mimic the stellar photosphere. This involved selecting the models with log(g) =5.0 and with stellar temperature within a wide range (2300-3500 K). We explored several stellar compositions by combining the metallicity [Fe/H] and the $\mathrm{\alpha}$ value of the star. In particular, we tested the metallicity [Fe/H] within the range [-1.0,1.0] with a 0.5 step. Then, we also evaluate the $\mathrm \alpha$ enrichment for the low metallicities cases: [Fe/H] = -0.5 dex with $\mathrm{ \alpha=+0.2 }$ and [Fe/H] = -1.0 dex with $\mathrm{\alpha=+0.4}$. 

The resulting spectra were multiplied for the HST G141 grism sensitivity \citep[calibrated by][]{2011wfc..rept....5K}, accounting for the conversion from flux to electrons units. Two other chromatic components were taken into account:  the dispersion of light from the star at the focal plane of the detector due to the grism and the dependence of the PSF (Point Spread Function) on the wavelenght observed. A detailed description of these two effects can be found in \cite{2017ApJS..231...13V}. Finally, the spectra are binned by using the same binning used to fit the spectral light curves.

Figures \ref{Met1} and \ref{Met2} show a comparison between the data and the various tested models, highlighting how the parameters (temperature, metallicity [Fe/H] and $\mathrm \alpha$ enrichment) impact the shape of the stellar spectrum. Specifically, we observe an increase in the amplitude of the features around $\mathrm{\sim1.3}$ $\mathrm{ \mu m}$ at lower temperature, and the rise in flux in the red end of the spectrum at higher metallicity.
The residuals between the observed data and the models are shown at the bottom of each plot. Finally in Tab. \ref{CHIQUADRO} we show the $\mathrm{\chi^{2}}$ values of the models at 3100, 3200 and 3300 K, that better fit the observations. Each $\mathrm{\chi^{2}}$ value is obtained using Eq. \ref{EQCHI} where N is the number of spectral bin observed, "$\mathrm{obs_{i}}$" are the observed spectral fluxes and $\mathrm{S_{i}}$ are the theoretical spectral fluxes predicted by the model for the specific stellar composition tested :
\begin{equation}
     \chi =\mathrm{ \sum_{i=1}^{N} \dfrac{(obs_{i}-S_{i})^{2}}{S_{i}}}
	\label{EQCHI}
\end{equation}

\begin{table}
\centering
\caption{List of $\mathrm{\chi^2}$ values for each stellar composition test with a given stellar temperature of 3100 K, 3200 K and 3300 K, the values of the stellar temperature nearest to the best fit of the data. "Z" is the metallicity and "$\mathrm \alpha$" the $\mathrm \alpha$ enrichment of the star.}
\label{CHIQUADRO}
\begin{tabular}{| c | c | c | c | c |}
\hline
\hline
\textbf{Z} &\textbf{\Large{$\mathrm{\alpha}$}} &\Large{$\mathrm{\chi^2_{3100 K }}$} &\Large{$\mathrm{\chi^2_{3200 K }}$}&\Large{$\mathrm{\chi^2_{3300 K }}$}\\
\hline
\rule{0pt}{4ex}-1.0&0.0&$\mathrm{4.30*10^{-3}}$&$\mathrm{4.76*10^{-3}}$&$\mathrm{4.95*10^{-3}}$\\
\rule{0pt}{4ex}-0.5&0.0&$\mathrm{3.06*10^{-3}}$&$\mathrm{3.43*10^{-3}}$&$\mathrm{3.71*10^{-3}}$\\
\rule{0pt}{4ex} 0.0&0.0&$\mathrm{1.45*10^{-3}}$&$\mathrm{1.92*10^{-3}}$&$\mathrm{2.44*10^{-3}}$\\
\rule{0pt}{4ex}+0.5&0.0&$\mathrm{5.25*10^{-4}}$&$\mathrm{1.05*10^{-3}}$&$\mathrm{1.85*10^{-3}}$\\
\rule{0pt}{4ex}+1.0&0.0&$\mathrm{2.32*10^{-4}}$&$\mathrm{8.17*10^{-4}}$&$\mathrm{2.03*10^{-3}}$\\
\rule{0pt}{4ex}-0.5&+0.2&$\mathrm{2.28*10^{-4}}$&$\mathrm{1.39*10^{-4}}$&$\mathrm{3.67*10^{-4}}$\\
\rule{0pt}{4ex}-1.0&0.4&$\mathrm{4.41*10^{-4}}$&$\mathrm{7.26*10^{-5}}$&$\mathrm{1.32*10^{-4}}$\\
\hline
\end{tabular}
\end{table}
Based on these tests, we identified a model of a quiet star with a combination of temperature, metallicity, and $\mathrm{\alpha \:(T = 3200 \: K, [Fe/H] = -1.0 \: dex, \alpha=+0.4)}$ that provides a good representation of the observed out-of-transit spectrum without evidence of chromaticity effects in the residuals, as illustrated in Figure \ref{Good_fit}. Note that we are comparing the data with a grid of models and not fitting the best temperature. This result, together with other previous evidences of the low level of activity of the star both from XUV observations \citep[see][]{2023AJ....165..200S} and from TESS almost flat light curves analysis for this target \citep{2022AAS...24030903B} and the good correspondence between the temperature of this model (3200 K) and that derived by literature \citep[$\mathrm{3216 \pm 39 \: K}$, derived by ][]{2019AJ....157...32M}, strongly suggests to rule out any strong stellar activity contribution in this transit observation. This lack of stellar activity implies that the planetary spectrum does not need a correction for stellar activity contaminations and that its spectral features are only due to the planetary atmosphere. 

\subsection{Stellar Activity Evaluation}
As a further check, we compared the observations with the simulated spectrum of a star with a dominant spot, following the methodology described in \cite{2021MNRAS.507.6118C}. \\
The star is modeled as having one dominant circular spot with radius $\mathrm{R_{s}}$ (normalized to the stellar radius $\mathrm{R_{\star}}$) on the stellar surface, whose projection onto the stellar disk is an ellipse with eccentricity dependent on the fractional distance $\mathrm d$ between the spot center and the center of the stellar disk. In this case the stellar disk is divided into N = 10 000 annuli, each with radius $\mathrm{r_{i}}$ and width $\mathrm{d_{r}}$ , and characterized by an emission $\mathrm{I_{\lambda}(r_{i})} $ given by a linear combination between the intensity emitted from the spot and the intensity emitted from photosphere. \\The out-of-transit flux $\mathrm{F_{\lambda}^{ out}} $ will be obtained by summing up the intensities from all the annuli:
\begin{equation*}
     \mathrm{F_{\lambda}^{ out} = \sum_{i=1}^N I_{\lambda}(r_{i})2\pi r_{i}d_{r} }
\end{equation*}
\begin{equation}
     \mathrm{I_{\lambda}(r_{i}) = ff_{i} \times I_{\lambda}(T_{s},r_{i}) + (1-ff_{i}) \times I_{\lambda}(T_{\star},r_{i})}
	\label{EQ3}
\end{equation}
where $\mathrm{ff_{i}}$ is the fraction of the $\mathrm{i^{th}}$ annulus covered by the spot. $\mathrm{I_{\lambda}(T_{s},r_{i})}$ and $\mathrm{I_{\lambda}(T_{\star},r_{i})}$ are the intensity profiles emitted from the photosphere and from the spot, respectively, for each annulus. Each intensity profile is obtained with a 4-coefficients limb darkening law \citep{2012A&A...546A..14C}:
\begin{equation}
     \mathrm{\dfrac{I_{\lambda}(\mu_{i}) } {I_{\lambda}(1) } = 1 - \sum_{i=1}^4 a_{n,\lambda} (1- \mu_{i}^{n/2})}
	\label{EQ4}
\end{equation}

with $\mathrm{\mu_{i} = \sqrt{1 - r_{i}^2}} $ and $\mathrm{a_{n,\lambda}} $ the four LDCs. Here, we neglected the dependence of the limb darkening effect on the temperature and we assumed the same LDCs \citep[the same from][]{2021AJ....161...44E} for the spots and the photosphere.
This model identifies the spot with 3 parameters: d, $\mathrm{R_{s}}$ and $\mathrm{T_{s}}$. Using as fitting algorithm the MCMC code emcee with the same methodology of \cite{2021MNRAS.507.6118C} we obtained as a result the fit showed in Figure \ref{Good_fit}.
Comparing the results of the model corresponding to a different composition of the star and the results of the model involving spot presence (Figure \ref{Good_fit}) it is clear that the 
 second one has a standard deviation much higher than the first one and a clear chromaticity of the residuals. Furthermore, the results of the fit accounting for the stellar activity suggest the presence of a big spot, with $\mathrm{R_{s} \sim 0.57 \: R_{\star}}$, results not compatible with the results of \cite{2023AJ....165..200S} which found a low level of activity for LHS 1140.\\
As a consequence we are quite confident that the spectral modulation studied by \cite{2021AJ....161...44E} is due to the planetary atmosphere without relevant stellar activity contamination. Our results also suggests that stellar composition must be considered in analysis of exoplanetary transmission spectra. 

\section{Stellar Population Membership}
Low metallicity values and $\mathrm \alpha$-element enrichment are typical of old stars of the thick disk \citep[compatible with the estimated age of this star $> 5$ billion years,][]{2019A&A...627A.144S}. We, therefore, analysed the kinematics of the target to determine its kinematics age. Galactic-spatial velocity components ($U$, $V$, $W$) were computed using Gaia DR3 \citep{2023A&A...674A...1G} proper motions, parallax and radial velocities by using the methodology described in \citet{2010A&A...521A..12M,2020A&A...644A..68M}.
The derived values are $U\mathrm{ = 4.38 \pm 0.04\: km \:s^{-1}}$, $V\mathrm{ =-49.52 \pm 0.09\: km\: s^{-1}}$ and $W\mathrm{ =4.2 \pm 0.4\: km \:s^{-1}}$, while the corresponding Toomre diagram is shown in Figure \ref{Toomre_LHS.png}.  To classify the star as belonging to the thin/thick disc population we made use of the procedure described by \citet{2003A&A...410..527B,2005A&A...433..185B}. We found that LHS1140 shows a kinematics compatible with an old thin disk star, populating the most external thin disk region of the Toomre diagram (Figure \ref{Toomre_LHS.png}). However, the kinematics criterion is a statistical one and alone is not enough to classify an individual star as a member of the thick or thin disk population. \\
In order to further clarify the chemical composition of LHS 1140, we analysed the available HARPS high-resolution optical spectra of this star. We use the methodology developed in \citet{2020A&A...644A..68M} which is based on the use of principal component analysis and sparse Bayesian methods and is calibrated using M dwarfs in binary systems around an FGK primary. We derived a metallicity value of [Fe/H] = -0.38 dex, thus, confirming that LHS 1140 is a metal poor star. Moreover, we find that $\mathrm \alpha$ elements show higher abundances ([Mg/H] = 0.01 dex, [Si/H] = -0.15 dex, [Ca/H] = -0.10 dex).
We should caution to use quantitative results from \citet{2020A&A...644A..68M} since it might not be optimal for metal poor stars, as the number of "training" stars in this region of metallicities is rather low.
However also this analysis points toward a poor metal star with enhanced $\mathrm{\alpha}$ elements, that cannot be properly described by assuming  solar-like atmospheric composition models. 

\begin{table}
\centering
\caption{\label{PRIORS} Priors of Taurex atmospheric retrieval framework used in our work. "$\mathrm{V_{x}}$" is the volume mixing ratio of the molecular species studied.}
\begin{tabular}{l|c|c}
\hline
\hline
\rule{0pt}{4ex} \textbf{Parameter} & \textbf{Prior Bounds} & \textbf{Scale}\\
\hline
\rule{0pt}{4ex} $\mathrm{V_{x}}$& -12, 0 &$\mathrm{log_{10}}$\\
\rule{0pt}{4ex} $\mathrm{T_{eq} (K)}$& 50, 500 &linear\\
\rule{0pt}{4ex} $\mathrm{P_{clouds} (Pa)}$&-4, 6&$\mathrm{log_{10}}$\\
\rule{0pt}{4ex} $\mathrm{R_{p} (R_{jup})}$&0.123, 0.185&linear\\
\rule{0pt}{4ex} $\mathrm{He/H_{2}}$&0.172&fixed\\
\rule{0pt}{4ex} $\mathrm{N_{2}/H_{2}}$&0.2&fixed\\
\hline
\end{tabular}
\end{table}

\section{Atmospheric Retrieval}
Given our analysis of the chemical composition of LHS 1140 and the consequent attribution of the features of the in-transit spectrum to the planet, we reviewed the work of \cite{2021AJ....161...44E} analyzing the planetary spectrum by using the Taurex retrieval framework \citep{2021ApJ...917...37A} with the same priors of \cite{2021AJ....161...44E} (shown in Tab. \ref{PRIORS}) and an expanded grid of atmospheric parameters with different combinations of $\mathrm{H_{2}O}$, $\mathrm{CH_{4}}$, $\mathrm{ N_{2}}$ and clouds as in Table \ref{MODELS}. 
Our aim was to confirm the results of \cite{2021AJ....161...44E} using the updated values for the planetary mass and radius from \cite{2023arXiv231015490C}, because the mass in particular can have a relevant impact in atmospheric retrievals \citep{2023A&A...669A.150D}.
We wanted also to include the flat case of an atmosphere without any relevant molecular contribution in the planetary spectrum for reference.
In order to efficiently explore the parameter space, TauREx uses the MultiNest optimizer \citep{2009MNRAS.398.1601F,2014A&A...564A.125B} and we chose to use 3000 live points to achieve a higher precision for our fits.\\
To efficiently explore the parameter space, we used TauREx with the MultiNest optimizer, employing 3000 live points to enhance fitting precision. In our retrieval procedure we used the molecular cross sections of $\mathrm{H_{2}O}$ and $\mathrm{CH_{4}}$ taken from \cite{2018MNRAS.480.2597P} and \cite{2017A&A...605A..95Y} respectively. Finally we took into account collision-induced-absorption (CIA) due to $\mathrm{H_{2}-H_{2}}$ 
\citep{2011JPCA..115.6805A,2018ApJS..235...24F} and to $\mathrm{H_{2}-He}$ and also Rayleigh scattering for alle the molecules in the atmosphere. \\
Following the criterion illustrated in \cite{doi:10.1080/01621459.1995.10476572}, bayesian evidence does not allow us to identify any specific atmospheric composition, probably because of the wavelength range and low resolution of our data that do not allow us to distinguish the bands of the specific molecules. However, the bayesian evidence suggests that the hypothesis of a flat atmosphere can be rejected. Furthermore it led to a slightly higher probability for $\mathrm{H_{2}O}$ comprehensive scenarios (Table \ref{MODELS}) with respect to $\mathrm{CH_{4}}$ rich atmosphere. The corner plot corresponding to the best fit is shown in Figure \ref{POSTERIOR}.
Figure \ref{DOMINATED} instead shows the comparison between the observed HST data and the two most extreme scenarios that we simulated: H$_{\rm 2}$O dominated and CH$_{\rm 4}$-dominated atmosphere, without cloud contributions. It is clear from the figure that, despite some small differences, both curves are similar and hard to distinguish.

\section{Conclusions}
LHS 1140 b is a planet close to the HZ of its host star. Previous studies \citep[e.g.][]{2021AJ....161...44E} analyzed its atmosphere suggesting the presence of water, but could not rule out the possibility that the modulation of the in-transit spectrum could be due to stellar activity contamination.\\
In this work, we found that the out-of-transit spectra of LHS 1140 during HST observations can be explained by a quiet, inactive star, with low [Fe/H] and high $\mathrm{ \alpha}$ abundances, so the difference between the in- and -out transit spectra can be entirely attributed to the LHS 1140 b planet atmosphere. Therefore, we are confident that the derived planetary spectrum 
is not contaminated by a significant stellar contribution and that any spectral feature presented does indeed reflects the planetary atmosphere of LHS 1140 b.
A proper evaluation of the atmospheric composition of LHS 1140 b will therefore require additional observations, as example with JWST or the upcoming Ariel mission, with higher spectral resolution or broader band coverage. In particular, we simulated the atmospheric signal of LHS 1140 b during a transit as observed with ARIEL tier 2, obtained through ARIELRAD \citep{2019EPSC...13..270M} and we did the same also for JWST Nirspec \citep{2022A&A...661A..80J} with G140M grism, using PANDEXO \citep{2017PASP..129f4501B} to simulate the expected observed flux for this target. Figure \ref{ARIEL} shows the results of this simulations, that is, the comparison between the simulated  atmosphere of LHS 1140 b with the same scenarios analyzed in Figure \ref{DOMINATED}. It can be seen that the difference between both cases is quite clear, and therefore detectable with ARIEL or JWST observations, mainly thanks to the long wavelength coverage, thus, allowing us to confirm or rule out the presence of water on the planet and to retrieve its abundance. The challenges in properly modeling the atmosphere of LHS 1140 b using HST WFC3 observations may arise from their low resolution and narrow spectral range together with a low signal to noise. 
We finally stress the importance of an accurate modelling of the host stellar spectra, not limited to stellar activity but including also its chemical composition, when dealing with suspect features in the retrieved in-transit spectra.

\begin{table*}
\centering
\caption{\label{MODELS} Fit corresponding to the many models tried to fit the planetary spectrum. We use a combination of $\mathrm{H_{2}O}$, $\mathrm{CH_{4}}$ as active molecules, with and without clouds and scenarios where we impose that $\mathrm{H_{2}O}$ or $\mathrm{CH_{4}}$ are above $\mathrm 10\%$ of the planetary atmospheres ("$\mathrm{H_{2}O}$ dominated" and $\mathrm{CH_{4}}$ dominated"). We also show the results obtained from a flat model. $\mathrm{Log}\mathcal{Z}$ is the bayesian evidence for each model. T is the temperature retrieved for the planet and $\mathrm{ R_{p}}$ its radius. When a parameter is not used in the model, it is labeled as "-". When not specified the atmospheric fill gases are H and He with a ratio of 0.172, while in other cases $\mathrm{N_{2}}$ is also present as fill gas and its concentration is fitted. }
\begin{tabular}{| c | c | c | c | c | c | c | c |}
\hline
\hline
\textbf{MODEL} &\textbf{log $\mathcal{Z}$} &\textbf{$\mathrm{ R_{p}( R_{J})}$ } &\textbf{ T (K) } &\textbf{ log($\mathrm{H_{2}O}$)} &\textbf{ log($\mathrm{CH_{4}}$) } &\textbf{ log($\mathrm{N_{2}/H_{2}}$)} &\textbf{log $\mathrm{P clouds}$}\\
\hline
\rule{0pt}{4ex} $\mathrm{N_{2}-H_{2}O}$ no clouds & 195.0 & $ \mathrm{ 0.1548^{+0.0006}_{-0.0007}}$ &  $ \mathrm{ 289^{+133}_{-121}}$ &  $ \mathrm{ -0.5^{+0.3}_{-0.3}}$ &  - &  $ \mathrm{ -3^{+3}_{-3}}$ &  - \\
\hline
\rule{0pt}{4ex} $\mathrm{N_{2}-H_{2}O}$-clouds & 194.4 & $ \mathrm{ 0.1540^{+0.0009}_{-0.0016}}$ &  $ \mathrm{ 312^{+111}_{-122}}$ &  $ \mathrm{ -0.5^{+0.3}_{-0.3}}$ &  - &  $ \mathrm{ -4^{+3}_{-2}}$ &  $ \mathrm{ 3.6^{+1.4}_{-1.3}}$ \\
\hline
\rule{0pt}{4ex} $\mathrm{N_{2}-H_{2}O-CH_{4}}$ no clouds & 193.9 & $ \mathrm{ 0.1547^{+0.0006}_{-0.0008}}$ &  $ \mathrm{ 276^{+127}_{-113}}$ &  $ \mathrm{ -0.6^{+0.3}_{-0.3}}$ &  $ \mathrm{ -0.6^{+0.3}_{-0.3}}$ &  $ \mathrm{ -3^{+4}_{-4}}$ &  - \\
\hline
\rule{0pt}{4ex} $\mathrm{N_{2}-H_{2}O-CH_{4}}$-clouds & 193.6 & $ \mathrm{ 0.1544^{+0.0007}_{-0.0011}}$ &  $ \mathrm{ 285^{+126}_{-134}}$ &  $ \mathrm{ -0.6^{+0.3}_{-0.3}}$ &  $ \mathrm{ -0.6^{+0.3}_{-0.3}}$ &  $ \mathrm{ -3^{+4}_{-3}}$ &  $ \mathrm{ 3^{+2}_{-3}}$ \\
\hline
\rule{0pt}{4ex} $\mathrm{N_{2}-CH_{4}}$-clouds & 192.7 & $ \mathrm{ 0.153^{+0.001}_{-0.005}}$ &  $ \mathrm{ 246^{+163}_{-112}}$ &  - &  $ \mathrm{ -5^{+3}_{-5}}$ &  $ \mathrm{ -2^{+.3}_{-5}}$ &  $ \mathrm{ 2^{+2}_{-5}}$ \\
\hline
\rule{0pt}{4ex} $\mathrm{N_{2}-CH_{4}}$ no clouds & 192.6 & $ \mathrm{ 0.1553^{+0.0006}_{-0.0009}}$ &  $ \mathrm{ 219^{+200}_{-91}}$ &  - &  $ \mathrm{ -4^{+2}_{-3}}$ &  $ \mathrm{ -1^{+2}_{-5}}$ &  - \\
\hline
\rule{0pt}{4ex} $\mathrm{H_{2}O \: dominated}$ no clouds & 191.7 & $ \mathrm{ 0.1559^{+0.0004}_{-0.0004}}$ &  $ \mathrm{ 134^{+57}_{-46}}$ &  $ \mathrm{ -0.6^{+0.4}_{-0.3}}$ &  - &  - &  - \\
\hline
\rule{0pt}{4ex} $\mathrm{H_{2}O}$ no clouds & 191.4 & $ \mathrm{ 0.1553^{+0.0005}_{-0.0005}}$ &  $ \mathrm{ 139^{+52}_{-47}}$ &  $ \mathrm{ -2.7^{+1.5}_{-1.2}}$ &  - &  - &  - \\
\hline
\rule{0pt}{4ex} $\mathrm{CH_{4} \: dominated}$-clouds & 191.1 & $ \mathrm{ 0.153^{+0.002}_{-0.003}}$ &  $ \mathrm{ 274^{+138}_{-136}}$ &  - &  $ \mathrm{ -0.5^{+0.4}_{-0.3}}$ &  - &  $ \mathrm{ 0.8^{+1.3}_{-1.9}}$ \\
\hline
\rule{0pt}{4ex} $\mathrm{H_{2}O \: dominated}$-clouds & 191.0 & $ \mathrm{ 0.1550^{+0.0006}_{-0.0019}}$ &  $ \mathrm{ 176^{+112}_{-62}}$ &  $ \mathrm{ -0.6^{+0.4}_{-0.3}}$ &  - &  - &  $ \mathrm{ 3^{+2}_{-5}}$ \\
\hline
\rule{0pt}{4ex} $\mathrm{H_{2}O}$-clouds & 191.0 & $ \mathrm{ 0.152^{+0.001}_{-0.009}}$ &  $ \mathrm{ 201^{+179}_{-71}}$ &  $ \mathrm{ -5^{+2}_{-5}}$ &  - &  - &  $ \mathrm{ 2^{+2}_{-5}}$ \\
\hline
\rule{0pt}{4ex} $\mathrm{CH_{4}}$-clouds & 191.0 & $ \mathrm{ 0.150^{+0.003}_{-0.005}}$ &  $ \mathrm{ 259^{+136}_{-129}}$ &  - &  $ \mathrm{ -5^{+2}_{-5}}$ &  - &  $ \mathrm{ 2^{+2}_{-4}}$ \\
\hline
\rule{0pt}{4ex} flat model & 190.4 & $ \mathrm{ 0.154^{+0.001}_{-0.003}}$ &  $ \mathrm{ 267^{+160}_{-145}}$ &  - &  - &  - &  - \\
\hline
\rule{0pt}{4ex} $\mathrm{CH_{4} \: dominated}$ no clouds & 189.9 & $ \mathrm{ 0.1552^{+0.0007}_{-0.0010}}$ &  $ \mathrm{ 226^{+185}_{-107}}$ &  - &  $ \mathrm{ -0.3^{+0.1}_{-0.3}}$ &  - &  - \\
\hline
\rule{0pt}{4ex} $\mathrm{CH_{4}}$ no clouds & 189.5 & $ \mathrm{ 0.1551^{+0.0007}_{-0.0007}}$ &  $ \mathrm{ 275^{+131}_{-121}}$ &  - &  $ \mathrm{ -0.2^{+0.1}_{-0.2}}$ &  - &  - \\
\hline

\end{tabular}

\end{table*}

\section*{Acknowledgements}

The authors acknowledge the support of the ASI-INAF agreement n. 2021-5-HH.0 and thank Angelos Tsiaras for his invaluable contribution in the processing and the reduction of HST WFC3 data, and Billy Edward for his helpful advices.\\
J. M. acknowledges support from the Italian Ministero dell'Università e della Ricerca and from the European Union - Next Generation EU through project PRIN MUR 2022PM4JLH ``Know your little neighbours: characterizing low-mass stars and planets in the Solar neighbourhood''.\\ 
Based on data products from observations made with ESO Telescopes at the La Silla Paranal Observatory under programms  ID 191.C-0873(A), 198.C-0838(A), 100.C-0884(A), 60.A-9709(G).
Part of the research activities described in this paper were carried out with contribution of the Next Generation EU funds within the National Recovery and Resilience Plan (PNRR), Mission 4 - Education and Research, Component 2 - From Research to Business (M4C2), Investment Line 3.1 - Strengthening and creation of Research Infrastructures, Project IR0000034 – “STILES - Strengthening the Italian Leadership in ELT and SKA”.

\section*{Data Availability}

The analysis shown in this paper is derived from publicly available data and codes: \url{https://archive.stsci.edu/hst/} (MAST portal), \url{https://github.com/ucl-exoplanets/Iraclis} \citep[Iraclis,][]{2018AJ....155..156T}, \url{https://phoenix.ens-lyon.fr/Grids/} \citep[Phoenix,][]{2012A&A...546A..14C}
\url{https://github.com/ucl-exoplanets/TauREx3_public} \citep[TauReX,][]{2021ApJ...917...37A}, \url{https://github.com/ucl-exoplanets/pylightcurve} \citep[pylightcurve,][]{2016ApJ...832..202T}, \url{https://emcee.readthedocs.io/en/stable/} \citep[emcee,][]{2013ascl.soft03002F}.\\ The specific result can be shared on reasonable request to the corresponding author.



\bibliographystyle{mnras}
\bibliography{MNRAS_MAIN} 




\appendix
\section{Pre and Post Transit Comparison}
\begin{figure}
		\centering
		\includegraphics[width=\columnwidth]{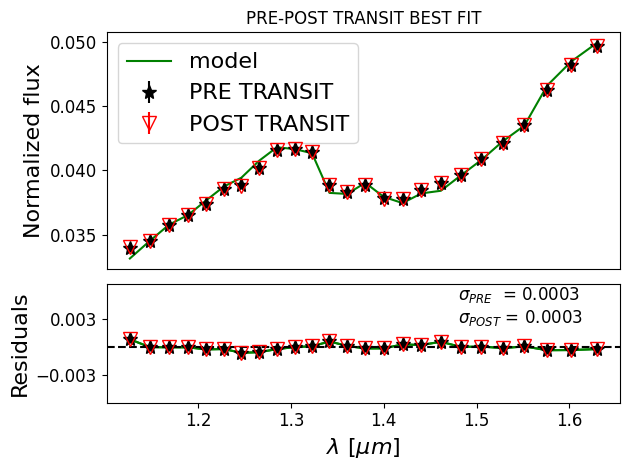}
		\caption{\label{PREPOST} Comparison between our best stellar model (green line), the pre-transit (black stars) and the post-transit (red triangles) mean stellar spectrum. It is clear that the pre-transit and post-transit spectra match each other with great precision and that their differences are negligible with respect to the residuals of the common best fit model. The error bars are too little to be visible in the figure. } 
\end{figure}

As discussed in section 3, processed data could be affected by time-dependent chromatic effects. For this reason we analyzed separately the pre-transit and post transit data 
to look for any relevant difference, possible clue of time-dependent systematics in our data.
In Fig. \ref{PREPOST} we show the resulting out-of-transit stellar spectra before and after transit. These two spectra clearly match each other and their differences are lower than their associated error and much lower than the differences between these spectra and their common best model.
As a further check, we performed the analysis using only pre or post transit spectra to verify the self-consistency of our results about the stellar composition of the star. The results obtained from those analyses (see Table \ref{CHIPRE}) are consistent with the results reported in Table \ref{CHIQUADRO} confirming the self-consistency of our approach and the temperature and metallicity found in section 4.1.

\begin{table}
\centering
\caption{ List of $\mathrm{\chi^2}$ values from the analysis of pre-transit/post-transit spectra. We show for each stellar composition test the $\mathrm{\chi^2}$ at a given stellar temperature of 3100 K, 3200 K and 3300 K, the values of the stellar temperature nearest to the best fit of the data.}
\label{CHIPRE}
\begin{tabular}{| c | c | c | c | c |}
\hline
\hline
\textbf{Z} &\textbf{\Large{$\mathrm{\alpha}$}} &\Large{$\mathrm{\chi^2_{3100 K }}$} &\Large{$\mathrm{\chi^2_{3200 K }}$}&\Large{$\mathrm{\chi^2_{3300 K }}$}\\
\rule{0pt}{3ex}&&pre/post&pre/post&pre/post\\
\hline
\rule{0pt}{4ex}-1.0&0.0&$\mathrm{(4.29/4.29)*10^{-3}}$&$\mathrm{(4.75/4.75)*10^{-3}}$&$\mathrm{(4.94/4.94)*10^{-3}}$\\
\rule{0pt}{4ex}-0.5&0.0&$\mathrm{(3.05/3.05)*10^{-3}}$&$\mathrm{(3.42/3.42)*10^{-3}}$&$\mathrm{(3.70/3.70)*10^{-3}}$\\
\rule{0pt}{4ex} 0.0&0.0&$\mathrm{(1.45/1.45)*10^{-3}}$&$\mathrm{(1.91/1.91)*10^{-3}}$&$\mathrm{(2.44/2.44)*10^{-3}}$\\
\rule{0pt}{4ex}+0.5&0.0&$\mathrm{(5.20/5.22)*10^{-4}}$&$\mathrm{(1.04/1.04)*10^{-3}}$&$\mathrm{(1.85/1.85)*10^{-3}}$\\
\rule{0pt}{4ex}+1.0&0.0&$\mathrm{(2.29/2.30)*10^{-4}}$&$\mathrm{(8.12/8.14)*10^{-4}}$&$\mathrm{(2.02/2.02)*10^{-3}}$\\
\rule{0pt}{4ex}-0.5&+0.2&$\mathrm{(2.25/2.26)*10^{-4}}$&$\mathrm{(1.37/1.38)*10^{-4}}$&$\mathrm{(3.65/3.66)*10^{-4}}$\\
\rule{0pt}{4ex}-1.0&0.4&$\mathrm{(4.39/4.40)*10^{-4}}$&$\mathrm{(7.16/7.26)*10^{-5}}$&$\mathrm{(1.32/1.33)*10^{-4}}$\\
\hline
\end{tabular}
\end{table}

\bsp	
\label{lastpage}
\end{document}